\documentclass[onecolumn,showpacs,amsmath,amssymb,prd,nofootinbib]{revtex4}
\usepackage{epsfig}
\def\Journal#1#2#3#4{{#1} {\bf #2}, #3 (#4)}

\def\NPB{{\em Nucl. Phys.} B}

\def\PRD{{\em Phys. Rev.} D}

\def\be{\begin{equation}}
\def\ee{\end{equation}}
\def\bea{\begin{eqnarray}}
\def\eea{\end{eqnarray}}

\begin{document}
%\vspace*{4cm}
\title{MOND and its bimetric formulation}
\author{Mordehai Milgrom}
\affiliation{Department of Particle Physics and Astrophysics, Weizmann Institute,\\Rehovot 76100, Israel}

\begin{abstract}
I first give a succinct account of the MOND paradigm--emphasizing the centrality of scale invariance in the nonrelativistic, deep-MOND limit--and describing rudiments of its phenomenology. I then present my credo, and some generalities, concerning existing MOND theories. Then I concentrate on one relativistic formulation of MOND in the form of a bimetric theory (BIMOND). I describe its various limits: the weak field, with application to gravitational waves, the nonrelativistic limit, and their further deep-MOND (low acceleration) limits, which are scale invariant. Other aspects of BIMOND that have been explored are aspects of cosmology, matter fluctuations in cosmology, and matter-twin-matter interactions. BIMOND remains largely unexplored, despite its promise in several regards: It tends to GR for $a_0\rightarrow 0$ ($a_0$ is the MOND constant); it has a simple nonrelativistic limit; it describes gravitational lensing correctly; and, it has a generic appearance of a cosmological-constant term that is of order $a_0^2/c^4$, as observed.
\end{abstract}
\maketitle
\section{The MOND paradigm}
The MOND paradigm of modified dynamics \cite{milgrom83} departs from
Newtonian dynamics and general relativity (GR) in the limit of small accelerations. Its goal is to account for all the mass discrepancies in the universe without invoking dark matter (and `dark energy'). Reference \cite{fm12} is an extensive recent review of MOND. The paradigm introduces a constant, $a_0$, with the dimensions of accelerations, at and below which MOND departs from standard dynamics. In the relativistic context it may be  useful, instead, to use the MOND length,\footnote{However, the threshold for galactic phenomena is defined by an acceleration, $a_0$, not by a length, even though these phenomena are nonrelativistic.} $\ell_{M}\equiv c^2/a_0$. For much higher accelerations--formally represented by taking $a_0\rightarrow 0$ in a MOND theory--we require restoration of Newtonian dynamics and GR. Another central tenet of MOND is that in the low-acceleration, or deep-MOND limit (DML), the equations of motion of nonrelativistic (NR) MOND are space-time scale invariant (SI) for purely gravitational systems \cite{milgrom09a}.\footnote{The action is not necessarily SI, but multiplied by a constant under dilatations.}$^,$\footnote{It is not clear whether this symmetry has a deep origin, or just happens to be a symmetry of the DML, on a par with the scaling symmetry $(t,{\bf r})\rightarrow(\lambda t,\lambda^{2/3}{\bf r})$ enjoyed by Newtonian dynamics. Either way the symmetry is powerfully predictive.}
\par
The DML of a MOND theory may be defined by applying a dilatation $(t,{\bf r})\rightarrow\lambda(t,{\bf r})$ to the degrees of freedom (DoF) in the equations of motion, and letting $\lambda\rightarrow\infty$. In this way, all DoFs with dimensions of acceleration tend to 0, as befits the DML. Also, if the limit exists it is automatically SI (because further, finite dilatations have no effect). If in addition the limit is nontrivial, in the sense that it describes the physics we want, the theory qualifies as a candidate MOND theory. For a pure gravity theory that involves as constants only $a_0,~G,~c$, and masses, an equivalent road to the DML is to take the limit $a_0\rightarrow\infty$, $G\rightarrow 0$, with $\Omega_0\equiv Ga_0$ fixed.\footnote{Seen by accompanying the dilatation by the same scaling of the length and time units. The two together leave the values of the DoFs intact, but change the values of the constants: $a_0\rightarrow \lambda a_0$, $G\rightarrow\lambda^{-1}G$.} In a nontrivial DML, only $c,~\Omega_0$ and the masses remain.\footnote{$\Omega_0$ has dimensions $[l]^4/[m][t]^4$; so, like $c$ and $m_i$ does not change under rescaling of the time and length units by the same factor.} If other dimensioned constants are involved they would have to go to their own limits. For example, a MOND extension of electromagnetism has not been considered to my knowledge. But, in such a context we would have to have charges (and currents) $e\rightarrow \lambda^{-1/2}e\rightarrow 0$, such that $e^2 a_0$ remain fixed in the limit, and electromagnetic interactions remain finite in a SI theory.
\par
Even without invoking a concrete MOND theory, the above basic tenets (with some additional implicit assumptions) lead to a large number of predictions--in the form of ``MOND laws''--for galaxies and galactic systems \cite{milgrom12a}.
These laws underlie predictions of observed regularities, such as the Tully-Fisher and Faber-Jackson relations, the appearance in different, independent contexts of a special surface-density constant ($\sim a_0/G$), and others. A robust prediction is that the asymptotic circular rotational speed around a mass $M$, $V_{\infty}$, is constant, and that $V_{\infty}\propto (MGa_0)^{1/4}$; $a_0$ is normalized such that $V_{\infty}=(MGa_0)^{1/4}$.
These laws also account for the dynamics within galaxies (including dwarf spheroidal galaxies) and groups of galaxies. In particular, the far field of isolated masses follows in full from the basic tenets.
Even all the salient aspects of rotation curves follow from these tenets.
All these {\it a priori} predictions of the MOND paradigm have been amply tested, and found to hold well, as described e.g. in Ref. \cite{fm12}. More recently published tests of these predictions concern the fields of two ellipticals to very large radii \cite{milgrom12b}, weak lensing in a large galaxy sample \cite{milgrom13}, internal dynamics of the dwarf satellites of the Andromeda galaxy \cite{mm13,mm13a}, polar-ring galaxies \cite{lueghausen13}, dynamics of the local group \cite{zhao13}, and a shell elliptical \cite{bilek13}.
\par
Two important challenges still remain for MOND. MOND explains away most of the mass discrepancy in galaxy clusters, reducing it from a factor $\sim 10$ in Newtonian dynamics to a factor of $\sim 2$. However, this factor 2 discrepancy remains unaccounted for by directly observed matter. It has been suggested that it is due to the presence of neutrinos or yet-undetected baryonic matter, but this remains to be vindicated. The challenge for MOND posed by the ``bullet'' cluster is not an independent new challenge, but part and parcel of this long known discrepancy.
The other major phenomenological task for MOND is to come up with a theory (which we do not yet have) that accounts in detail for cosmology and structure formation.
\par
Despite these challenges, the fact that MOND achieves so much, and does it so well, signifies to me that it is basically a correct paradigm, and that the challenges are only apparent, temporary obstacles.
\par
While many of the salient predictions of MOND follow from the basic tenets alone, and so are shared by all MOND theories that incorporate these tenets, these axioms alone do not predict many of the details of galactic phenomenology, such as the detailed gravitational field of a given mass distribution.\footnote{ This is similar to have had predictions based on the equivalence principle without GR, such as gravitational redshift and light bending, or the predictions of quantum mechanic made without a complete theory, such as black body spectrum, photoelectric effect, specific heat of solids, spectral series of H-like atoms, etc.} For these we need a full-fledged, consistent MOND theory to cover both NR and relativistic phenomena. Finding such a theory is particularly pressing in the context of cosmology.
\subsection{The significance of $a_0$}
The MOND constant $a_0$ appears in several of the MOND laws, as well as in more detailed predictions, such as full rotation curves of galaxies, and, so, can be determined in various {\it independent} ways from the data, all of which give consistently $a_0\approx (1.2\pm 0.2)\times 10^{-8}{\rm cm/sec^2}$. It was noticed early on \cite{milgrom83,milgrom89,milgrom94} that this value is of the order of cosmologically relevant accelerations:
We have $\bar a_0\equiv 2\pi a_0\approx cH_0\approx c^2(\Lambda/3)^{1/2}$ ($H_0$ is the Hubble constant, and $\Lambda$ the cosmological constant). Or, in other words, the MOND length is of order of today's Hubble distance:
$\ell_M\approx 2\pi \ell_H$ ($\ell_H\equiv c/H_0$), or of the de Sitter radius associated with $\Lambda$: $\ell_M\approx 2\pi \ell_S$. The MOND mass
$M_M=c^4\Omega_0^{-1}$ is also a useful reference mass, and is of order of the closure mass (or the total ``dark energy'') within today's horizon (or the de Sitter radius).
\par
This ``coincidence'' may be an important hint for understanding the origin of MOND, and constructing MOND theories (see below).
It also has several interesting consequences. Some examples:

(i) If a system of mass $M$, and size $R< \ell_H$, produces gravitational accelerations $MG/R^2< a_0$, then $MG/R< c^2/2\pi$: Namely, no system smaller than today's cosmological horizon requires for its description a strong-field ($MG/R\sim c^2$) deep-MOND description. Since MOND is probably a derived, effective paradigm, it is not clear that an effective MOND theory needs to have a consistent relativistic DML.

(ii) Strong lensing of cosmological sources (such as quasars) by a much nearer lens cannot prob the (deep) MOND regime.

(iii) Cherenkov losses by high-energy particles to subluminal gravitational waves, which may occur in some MOND theories, occur over distances of order $\ell_M$; so losses are unimportant for sub-Hubble travel \cite{milgrom11a}.
\section{\label{theories}MOND theories}
We work today with several NR, and several relativistic, full-fledged MOND theories to be listed below. They are all metric theories (or potential theories) that are, conceptually, straightforward generalizations of GR. All are derived from an action; all are modified-gravity theories\footnote{I.e., they can be written such that the standard matter action remains intact, and only the gravitational action is modified: the Poisson action in the NR case, and the Einstein-Hilbert action in the relativistic case.}; all involve $a_0$ as a constant in addition to $G$ (and $c$); all have a NR version that satisfies the basic premises of MOND; and, they all involve an interpolating function that has to be put in by hand, to artificially interpolate between the MOND and the high-acceleration regimes.
Constructing full-fledged theories on these principles has proven the easiest approach.
\par
However, past experience, with relativity and quantum mechanics, has demonstrated that the appearance of new regimes of physical phenomena, separated from the old ones by newly introduced constants ($c$ and $\hbar$) is not accompanied by some artificial interpolating function in the foundations of the theory (the action). Furthermore, different phenomena involve different ``interpolating functions'' (such as the Lorentz factor, or the particle dispersion relation in relativity, and the black-body function, or the specific heat of solids as a function of temperature, in quantum theory). And, these functions are derived within an all-encompassing theory. I expect a similar state of things in an eventual basic MOND theory.
Another warning sign comes from the above-mentioned coincidence of $\bar a_0\approx c^2(\Lambda/3)^{1/2}$ \cite{milgrom09a}.
\par
Thus, I feel that all the existing MOND theories are anything but the final word. At best, each theory, if relevant at all, can be considered only an effective, approximate theory, of limited validity.
\par
Still, since the advent of MOND such theories have proven very useful: They have provided proofs of various concepts, first that MOND theories can be written that are derived from an action, satisfy all the standard conservation laws, give the correct center-of-mass motion of a composite body, etc. Then, that covariant theories can be written with, e.g., correct lensing.
Thus, we should view these theories more as scaffolds, or as crutches for MOND to walk with until it can walk on its own legs. We use them for calculations, since they are complete theories, in the hope that they are close enough approximations for the problem at hand. This hope is founded on the fact that they all satisfy the basic tenets, and so all make the same salient predictions \cite{milgrom12a}. Further confidence in them is gained by noting that they also make very similar, if not always quite the same, predictions of more detailed aspects such as full rotation curves of spiral discs \cite{brada95,angus12,milgrom12c}, or of gravitational lensing.
\par
But, it is difficult to know what their exact validity domain is, and how far to trust their predictions beyond those that are anchored in the basic tenets. They may differ greatly among themselves, and from the ``correct'' MOND theory, on other, more subtle issues, such as small effects in the solar system, or the nature of the external-field effect \cite{milgrom11b}. There are many MOND theories that one can write, and the salient observations of galactic dynamics can hardly distinguish between them.
\par
Bimetric MOND (BIMOND) \cite{milgrom09}, our subject here, should also be viewed in this light.
\par
Many ideas have already been suggested that depart from the above scheme of constructing MOND theories. Some are more advanced, some less, but none have yet lead to a full-fledged theory. No doubt, fresh minds will come up with yet more ideas.
\par
There are, e.g., quite a few ideas to obtain MOND phenomenology from ``microscopic'' (quantum) approaches \cite{milgrom99,pikhitsa10,klinkhamer11,pa12,pazy13}, from some sort of an omnipresent medium \cite{bt08,zl10,ho12}, or from other constructs \cite{chang08}.
\par
Certainly, one should consider also MOND theories where the matter actions are modified [termed generally modified inertia (MI)].\footnote{The distinction is sometimes semantic; e.g. Brans Dicke or TeVeS can be formulated as either.} This approach has proven more recalcitrant, but the fact that no full-fledged MI has been constructed (there are some useful toy theories \cite{milgrom94,milgrom11b}) does not make this avenue less attractive and promising. More generally, we do not even know whether MOND has relevance only for gravitational phenomena, or whether it pertains to all dynamical phenomena.
After all, the Einsten-Hilbert action is the kinetic action of the gravitational DoFs; so why should the underlying phenomena leading to MOND affect this action and not other kinetic actions.
Most of the heuristic ideas to base MOND on microscopic phenomena \cite{milgrom99,pikhitsa10,klinkhamer11,pa12,pazy13}, are of this type.
\par
For example, the interaction of a body with the quantum vacuum in a non-flat universe, such as a de Sitter space time (as, e.g., encoded partly in the Unruh effect), may lead to non-standard inertia that depends also on the characteristic acceleration (or more general attributes) associated with the universe at large
($c^2/\ell_S$, in the case of a de Sitter universe) \cite{milgrom99}.
\par
Since BIMOND is the main subject here, I concentrate in what follows only on MG theories that are relevant for BIMOND. More details on MI theories can be found in \cite{milgrom94,milgrom99,milgrom11b}.
\par
In summary, there is still much to unravel regarding MOND's origin and its theory, and we probably need a completely new concept to build on.

\subsection{Existing nonrelativistic theories}
There are two NR theories relevant for BIMOND (see below); both modify the Poisson action and field equation, for the gravitational potential, $\phi$. In one \cite{bm84}, the
Poisson equation is replaced by a nonlinear version
\begin{equation}\vec \nabla\cdot[\mu(|\vec\nabla\phi|/a_0)\vec\nabla\phi]= 4\pi G \rho.\label{eq:i} \end{equation}
Very interestingly, its DML, $\vec\nabla\cdot(|\vec\nabla\phi|\vec\nabla\phi)= 4\pi \Omega_0\rho$, is
invariant under space conformal transformations \cite{milgrom97}. The symmetry group of DML theory for the potential is thus the same as the isometry group of a 4-D de Sitter space-time, with possible deep implications \cite{milgrom09a}.
\par
The second theory, Quaslinear MOND (QUMOND) \cite{milgrom10a} (also derived from an action), involves two potentials whose field equations are
\begin{equation}\Delta\phi^N= 4\pi G \rho,~~~~~~~~~\Delta\phi=\vec \nabla\cdot[\nu(|\vec\nabla\phi^N|)\vec\nabla\phi^N],\label{eq:ii} \end{equation}
requiring solving only the linear Poisson equation twice. The DML of the potential action (with $\psi\equiv a_0\phi^N)$: $\Delta\psi=4\pi\Omega_0\rho,~\Delta\phi=\vec\nabla(|\nabla\psi|^{-1/2}\nabla\psi)$,
is space scale invariant.\footnote{Generally the case for any MG MOND theory, since the DML equations of motion of a MOND theory are space-time SI, those equations where time does not appear (in time derivatives) are invariant under spatial dilatations (seen by keeping the space scaling dimension of the various DoFs the same as their space-time scaling dimension).}
\par
There are, in fact, various generalizations possible. For example, for two-potential theories, a first-order Lagrangian can be written \cite{milgrom10a} $\mathcal{L}=\mathcal{L}_g+(1/2)\rho {\bf v}^2({\bf r})$, with $\mathcal{L}_g=-\rho\phi({\bf r})+\mathcal{L}_f[(\vec\nabla\phi)^2,(\vec\nabla\psi)^2,\vec\nabla\phi\cdot\vec\nabla\psi)]$
that embody the MOND tenets. They have a DML of the form
\begin{equation} \mathcal{L}_f=\Omega_0^{-1}\sum_{a,b} A_{ab}[(\vec\nabla\phi)^2]^{a+3/2}[(\vec\nabla\psi)^2]^{a+b(2-p)/2}(\vec\nabla\phi\cdot\vec\nabla\psi)^{b(p-1)-2a}, \label{eq:iiaa}\end{equation}
where $p$ is fixed for a given theory, the 3rd tenet is satisfied for any set of $a,~b$.
The dimensions of $\phi$ and $\psi$ are, respectively, $[l]^{2}[t]^{-2}$ and $[l]^{2-p}[t]^{2(p-1)}$; $A_{ab}$ are dimensionless.
The DML gravitational Lagrangian $L_g=\int \mathcal{L}_gd^3x$ is invariant under spatial dilatations (${\bf r}\rightarrow \lambda{\bf r}$). For any $p$,
the nonlinear Poisson theory is gotten with $a=b=0$, while QUMOND is the special case with $p=-1$ and two terms with $a=-3/2,~b=1$  and $a=-b=-3/2$.

\subsection{\label{reltheories}Relativistic theories}
Apart from BIMOND, to be described below, the full-fledged relativistic MG theories propounded to date are:
(i) TeVeS (Tensor-Vector-Scalar Gravity) (\cite{bekenstein04}, based on ideas in \cite{sanders97}), and reviewed in \cite{skordis09}. It was modified with some benefit by Galileon k-mouflage adaptations \cite{babichev11}.
(ii) MOND adaptations of Aether theories \cite{zlosnik07,sanders11,marsat11}. (iii)
Nonlocal metric MOND theories \cite{deffayet11}.
\par
While it may well be that none of these (including BIMOND) is a step in the right direction in accounting for MONDian dynamics, remembering the many years when MOND had lacked a sound relativistic formulation, and the denunciation that came with it, we must view these attempts as reassuring. They are also useful in the senses alluded to above.

\section{\label{bimond}BIMOND}

In BIMOND, gravity is described by two metrics, $g_{\mu\nu}$ and $\hat g_{\mu\nu}$. Its action is
\begin{equation} I=-\frac{c^2}{16\pi G}\int[ \beta g^{1/2} R +\alpha{\hat g}^{1/2}\hat R
 +2v_{g\hat g}\ell_{M}^{-2}\mathcal{M}]d^4x
 +I_M(g_{\mu\nu},\psi_i)+\hat I_M(\hat g_{\mu\nu},\chi_i).\label{eq:iii} \end{equation}
It is made up of the standard Einstein-Hilbert actions for each of the metrics (with possible strength adjustments $\beta,~\alpha$), and an interaction term, whose strength is naturally set by the MOND length, $\ell_{M}$; $v_{g\hat g}$ is a combined volume form.  $I_M$ and $\hat I_M$ are the matter actions for matter and twin matter (TM), which may, in principle, exist and couple to $\hat g_{\mu\nu}$, as matter couples to $g_{\mu\nu}$.\footnote{TM is not related to ``dark matter'', and is not required for MOND phenomenology.} The dimensionless interaction, $\mathcal{M}$, is a function of scalars built from the dimensionless relative-acceleration tensors:
\begin{equation}\ell_{M} C^{\alpha}_{\beta\gamma},~~~~~~{\rm where}~~~~~C^{\alpha}_{\beta\gamma}=\Gamma^{\alpha}_{\beta\gamma}
 -\hat\Gamma^{\alpha}_{\beta\gamma},\label{eq:iv} \end{equation}
well apt in the MOND context. Here, the MOND length is introduced in another role ($\Gamma^{\alpha}_{\beta\gamma}
,~\hat\Gamma^{\alpha}_{\beta\gamma}$ are the Levi-Civita connections).
In particular I found the following scalars very useful as variables in $\mathcal{M}$, and they are used below
  \begin{equation}\Upsilon= g^{\mu \nu} \Upsilon_{\mu\nu},~~~\hat\Upsilon= \hat g^{\mu\nu}\Upsilon_{\mu\nu},~~~~~~~~~~~~\Upsilon_{\mu\nu}\equiv
C^{\gamma}_{\mu\lambda}C^{\lambda}_{\nu\gamma} -C^{\gamma}_{\mu\nu}C^{\lambda}_{\lambda\gamma}.\label{eq:v} \end{equation} Hereafter I take $c=1$, and for concreteness use $\beta=\alpha=1$. Other values are also of interest; in particular $\alpha=-\beta$ is an exceptionally special case. There are reasons to believe \cite{milgrom10c}
that we need to have $\beta\approx 1$.
\par
Aspects of BIMOND that have been considered preliminarily are: matter-TM interactions \cite{milgrom10b}, matter fluctuations in cosmology \cite{milgrom10c}, aspects of cosmology  \cite{milgrom09,cz10}, and the weak-field limit and gravitational waves \cite{milgrom13a}. Still, BIMOND remains largely unexplored, despite its promise in several regards: It tends to GR for $a_0\rightarrow 0$; it has a simple and elegant nonrelativistic (NR) limit; it describes gravitational lensing correctly; and, it has a generic appearance of a cosmological-constant term that is of order $a_0^2/c^4$, which would account for the $a_0$ coincidence.

\par
\section{Limits}
\subsection{The high-acceleration limit}
BIMOND is required to tend fast enough to GR for $g_{\mu\nu}$ in the limit
 $a_0\rightarrow 0$. By taking
$\mathcal{M}\rightarrow \mathcal{M}(\infty)=const$, we get two decoupled copies  of GR with a cosmological constant
$\Lambda\sim \mathcal{M}(\infty)a_0^2$.
\par
This means that in very high acceleration systems such as the inner solar system, binary compact objects, etc. BIMOND departs only very little from general relativity.
\par
The theory may not be expandable in $a_0$ near $a_0=0$. Such expandability is a central assumption in the discussion of  Ref.  \cite{boulanger01} regarding multi-metric theories, which thus may not apply to BIMOND.
\par
Decoupling also occurs when the two metrics are equal: we then
get GR with $\Lambda\sim \mathcal{M}(0)a_0^2$.
\subsection{\label{wfl}The weak-field limit}
Write $g_{\mu\nu}=\eta_{\mu\nu}+ h_{\mu\nu},~\hat g_{\mu\nu}=\eta_{\mu\nu}+ \hat h_{\mu\nu}$. Then, to lowest order in $h_{\mu\nu},~ \hat h_{\mu\nu}\ll 1$--the weak-field limit (WFL) of such a background-symmetric system \cite{milgrom13a}--the theory splits into two sectors for two linear combinations, $h_{\mu\nu}^{\pm}=h_{\mu\nu}\pm\hat h_{\mu\nu}$.
The WFL action is  \cite{milgrom13a}
\begin{equation} I\approx -\int d^4x[E(h_{\mu\nu}^+)+ h_{\mu\nu}^+\mathcal{T}^{+\mu\nu} +4a_0^2\bar{\mathcal{M}}(q^-)+ h_{\mu\nu}^-\mathcal{T}^{-\mu\nu}],\label{eq:vii} \end{equation}
where $E( h_{\mu\nu}^+)$ is the (quadratic) WFL of the Einstein Lagrangian, $\mathcal{T}^{\pm\mu\nu}\equiv \mathcal{T}^{\mu\nu}\pm\hat{\mathcal{T}}^{\mu\nu}$, the respective energy-momentum tensors; $q^-$ is a scalar
quadratic in $h_{\mu\nu,\sigma}^-/a_0$ and $\bar{\mathcal{M}}( q^-)$ is some WFL interaction that depends on the WFL of $\mathcal{M}$. Note that $q^-$ (and $\bar{\mathcal{M}}$) is not assumed small in the WFL, even though it is quadratic in $h$; it is $\sim (a/a_0)^2$. The $h_{\mu\nu}^+$ sector is equivalent to the WFL of GR, with the gauge freedom, and has the same vacuum gravitational waves. The $h_{\mu\nu}^-$ sector is fully nonlinear even for the smallest values of $h_{\mu\nu}^-$, and is left with no gauge freedom.
\par
Despite the strong nonlinearity, an arbitrary pair of harmonic GR wave packets of $h_{\mu\nu}$ and $\hat h_{\mu\nu}$ moving in the same direction, is a solution of the (vacuum) BIMOND WFL  \cite{milgrom13a}.

\subsection{\label{dmlwfl}The deep-MOND weak-field limit}
In the DML of the WFL ($a_0\rightarrow \infty, ~\Omega_0$ fixed), $\bar{\mathcal{M}}( q^-)$ becomes homogeneous of degree 3 in $h_{\mu\nu}^-$.
For purely gravitational systems, the equations of motion (including the geodesic equations and the gauge conditions, which involve no dimensioned constants) become scale invariant. Then, $h_{\mu\nu}^-$ scale as $(M/M_M)^{1/2}$.
\par
It follows, for example, that the light bending angle is independent of the impact parameter, $b$, for any static, lensing mass, $M$, distributed well within $b$ (so it can be considered a point mass), and $b\gg r_M\equiv (MG/a_0)^{1/2}$ ($r_M$ is the MOND radius of the mass), so that the DML applies. This constant angle is $\propto (M/M_M)^{1/2}$, but the proportionality factor depends on details of the theory.

\subsection{The nonrelativistic limit}

For the choice of scalar argument $\Upsilon,~\bar\Upsilon$ (but not generally), in the NR limit (WFL, slow motions) there is a gauge for which, as in GR:
$g_{\mu\nu}=\eta_{\mu\nu}-2\phi\delta_{\mu\nu}$, and $\hat g_{\mu\nu}=\eta_{\mu\nu}-2\hat\phi\delta_{\mu\nu}$. Defining
$\tilde\phi=\phi+\hat\phi$ and $\bar\phi=(\phi-\hat\phi)/2$, the field equations are (for $\alpha=\beta=1$):
 \begin{equation}\Delta\tilde\phi= 4\pi G (\rho+\hat\rho),~~~~~~~~~~
 \vec \nabla\cdot\{\tilde\mu(|\vec\nabla\bar\phi|/a_0)\vec\nabla\bar\phi\}= 4\pi G (\rho-\hat\rho)\label{eq:axe} \end{equation}
\par
For $\alpha+\beta=0$ we get instead QUMOND as the NR limit.

\subsection{The deep-MOND nonrelativistic limit}
For static sources, the basic tenets (SI) dictate that asymptotically far from a mass $M$, $h_{\mu\nu}^-$ dominate and become logarithmic in the radius from $M$: $h_{\mu\nu}^-\propto (M/M_M)^{1/2}{\rm ln}(r)$. In particular, $a_0$ is normalized so that $h_{00}^-\approx -2(M/M_M)^{1/2}{\rm ln}(r)$.
For the choice of scalar argument $\Upsilon,~\bar\Upsilon$, the light-bending angle is $2\pi(M/M_M)^{1/2}$. For this case we have in the DML
\begin{equation}\phi=-\hat\phi=\bar\phi,~~~~~~\vec \nabla\cdot[|\vec\nabla\bar\phi|\vec\nabla\bar\phi]=4\pi \Omega_0(\rho-\hat\rho)\label{eq:x} \end{equation}
This is conformally invariant to spatial transformations.
\subsection{Matter-Twin matter interactions}
The interaction of mater and TM in the NR limit is interesting \cite{milgrom10b}. It follows from the field equations ~(\ref{eq:axe}), and ${\bf a}=-\vec\nabla\phi,~~\hat{\bf a}=-\vec\nabla\hat\phi$. Thus we see that: (i) There is no MOND for $\rho=\hat\rho$ (decoupling for matter-TM-symmetric systems). (ii) MOND acts in full without TM ($\hat\rho=0$). (iii) There is no interaction in the Newtonian regime (peculiar to our choice $\beta=1$)
(iv) In the DML, matter and TM behave in all senses as if they have opposite gravitational masses (passive and active); so that each attracts its kind, but they repel each other.
\par
Conformal invariance allows us to calculate the general two-body force in the DML: ${\bf F}=-{\frac{2}{3}}
 (\Omega_0)^{1/2}[(M\pm m)^{3/2}
 -M^{3/2}-m^{3/2}]\frac{\bf r}{r^2}$, where the plus is for the same type, and minus for different mass types.

\section{\label{cosmology}Cosmology and perturbation growth}
BIMOND, and, more generally MOND, has still to address properly the roles played by cosmological dark matter (and perhaps ``dark energy'') in the standard dynamics, and to predict in detail the CMB and structure formation.
\par
Because of the cosmological connotations of $a_0$ alluded to above (and other possible connections between local MOND dynamics and cosmology), I have always felt  that the understanding of MOND's origin and of its effects in cosmology will have to come together, within one framework. This contrasts with the state of things in standard dynamics, where we first invent a theory that accounts for small, local systems, and then apply it to the universe at large as if it were just another system.
But, as discussed in Sec. \ref{theories}, existing MOND theories may, at best, be effective theories that approximate deeper a theory, and that are restricted in applicability to the description of cosmologically small systems, such as galaxies.
Then clearly, we cannot expect such theories to describe cosmology in detail, as we cannot expect the effective-mass theory in solids to apply beyond its limitations, and describe the band structure, for example.
\par
Having said that, it has to be realized that MOND does have clear aspects that can play the role of cosmological DM. For example, since fluctuations growth (say, after matter-radiation decoupling) is in the DML, it occurs much faster than in standard dynamics, thus potentially obviating the role of DM in enhancing structure growth \cite{sanders01,nusser02,llinares08,angus13}.
Specifically, this is also the case in BIMOND (see \ref{fluctuations} below).
And so, it is useful to explore the cosmological implications of the existing theories, to see how far we can go with them, and what is still missing.
\par
Some cosmological models and other aspects of cosmology in BIMOND have been discussed \cite{milgrom09,milgrom10c,cz10}. But, these studies have explored only limited subclasses of BIMOND, and, in any event, are a still far cry from a full account of the cosmological data as we know it.
\subsection{Perturbation growth}\label{fluctuations}
The development of small fluctuations in matter and TM densities in an expanding universe have been studied \cite{milgrom10c} in the BIMOND subclass based on the $\Upsilon_{\mu\nu}$ tensor, for the restricted case of a globally matter-TM symmetric theory ($\alpha=\beta$) and universe (equal mean mass densities for all matter and TM components). With $\delta{\bf v}$ and $\delta\rho$ the velocity and density-perturbation fields for matter, and hatted quantities for TM, and defining
\begin{equation}{\bf u}=\delta{\bf v}+\delta\hat{\bf v},~~~\bar{\bf u}=\delta{\bf v}-\delta\hat{\bf v},~~~
 \epsilon=(\delta\rho+\delta\hat\rho)/\rho_b,~~~\bar \epsilon=(\delta\rho-\delta\hat\rho)/\rho_b,\label{eq:xii} \end{equation}
we have in comoving coordinates ($\rho_b$ is the background density for both, $a$ the scale factor, $v_s$ the mean speed of sound)
\begin{equation} \dot\epsilon+a^{-1}\nabla\cdot{\bf u}=0,~~~~~~~~~~~~~~~~~~~~ \dot{\bar \epsilon} +a^{-1}\nabla\cdot\bar{\bf u}=0\label{eq:xiv} \end{equation}
\begin{equation} \dot{\bf u}+(\dot a/a){\bf u}=
- a^{-1}\vec\nabla\delta\tilde\phi-v_s^2 a^{-1}\vec\nabla\epsilon,~~~~~~~~~~~~ \dot{\bar{\bf u}}+(\dot a/a)\bar{\bf u}=
-2a^{-1}\vec\nabla\bar\phi-v_s^2 a^{-1} \vec\nabla\bar\epsilon\label{eq:xv} \end{equation}
\begin{equation} \Delta\delta\tilde\phi= 4\pi G  a^2\rho_b\epsilon,~~~~~~~~~~~~~~~~~~~~~~~ \nabla(\tilde\mu|\vec\nabla\bar\phi/aa_0|\vec\nabla\bar\phi) = 4\pi G  a^2\rho_b\bar \epsilon. \label{eq:xvi} \end{equation}
We note some important features that may also be present in the more general case: (i) In line with the general WFL of BIMOND, there is complete decoupling between the two sectors, one describing the sum, and the other the difference, between the matter and the TM perturbations. (ii) The sum sector behaves exactly as in GR. (iii) The minus sector is MONDian in that the governing potential of the growth of density differences satisfies a MONDian nonlinear Poisson equation. Thus $\bar\epsilon$  can grow much faster that $\epsilon$, in which case, very quickly we have  $\delta\rho\approx -\delta\bar\rho$, with each growing much faster than in GR (matter perturmations grow not only by their own self attraction, but also through ``shepherding'' by neighboring TM). So, matter and TM become separated. This process continues even when the fluctuations are strong, in light of the above-mentioned matter-TM repulsion in the MOND regime. Matter and TM are thus expected to form mutually avoiding structures, with concentrations of one residing in the voids of the other.
(iv) The growth of perturbation can be nonlinear even for the weakest of perturbations, with coupling between nested perturbations on all scales when their magnitude puts them in the MOND regime. (v) The interplay between pressure and gravity is here very different from standard dynamics, as epitomized, e.g., by the Jeans criterion. Here, gravity can overcome pressure for any mass, for weak enough a perturbation. (vi) To be able to calculate the growth of perturbations, we need to know the initial conditions (say at matter-radiation decoupling) not only for matter (which can be deduced from the CMB), but also for TM, which we do not know.

\section{Discussion}
Like other known, full-fledged, MOND theories, BIMOND employs an interpolating, bimetric interaction function that has to be put in by hand. BIMOND can thus, at best, be an effective MOND theory that, even if applicable for a range of phenomena, would have to be understood at a deeper level.
\par
But, even so, it may be very useful in several ways: It increases the variety of relativistic MOND formulations, enabling us to explore a different direction in search for a deeper theory. In a sense it also augments the confidence that a satisfying, relativistic form of MOND (with, e.g., correct lensing) will be found: after the long absence of relativistic formulations prior to the advent of TeVeS, we now have several disparate forms propounded within just a few years. BIMOND introduces new elements--not present in other relativistic MOND formulations--which perhaps will turn out to be parts of a future theory. BIMOND also points to a way in which a cosmological-constant of the order of $a_0^2$ may naturally appear in a theory.
\par
Only a small part of the BIMOND class of theories have been considered in some detail. The sub-classes receiving most attention are those with $\pm\alpha=\beta=1$, and with the rather limited  choice of scalar arguments based on contractions of the $\Upsilon_{\mu\nu}$ tensor.\footnote{In one brief departure, it was shown \cite{milgrom09}, that using more general quadratic scalars
leads to a multi-potential NR limit, which is mathematically different from Eq.~(\ref{eq:axe}).}
\par
There are also important matter-of-principle issues yet to be checked, such as, causality, appearance of ghosts, and stability. A related issue concerns the study of gravitational waves in more detail (in other versions of the theory, on backgrounds, etc.).
\par
More generally, the bimetric structure of BIMOND calls for, and may point to, a deeper foundation of the theory. While having a space time with one geometry is quite natural, what is the meaning of two geometries on the same space time? Such a system, with two metrics but one set of gauge conditions is also generically beset by ghosts (this has not been checked in the case of BIMOND, which is not an ordinary bimetric theory).
While this is a question that may arise in any bimetric theory, it is
especially acute here, with the TM metric playing a symmetric role to the matter one, and with the potentiality of the existence of TM that couples only to its own geometry. Perhaps BIMOND is an effective, approximate theory to one that describes two interacting 4-D space times (``membranes'').
What then is the nature of the interaction, and how does it arise? What are the conditions for, and nature of, the one-to-one correspondence between events on the two manifolds that affords a single-manifold approximation, and the approximate locality of what must, otherwise, be a nonlocal interaction between the manifolds? And, how is one set of diffeomorphisms lost in the process?

\end{document}